\documentclass[12pt,a4paper,onesided]{article}

\usepackage[latin1]{inputenc}
\usepackage{graphicx,color}
\usepackage{epsfig}
\usepackage{verbatim}
\usepackage[german,english]{babel}
\usepackage{amsmath,amsthm,amssymb}
\usepackage[backend=bibtex,defernumbers=true,sorting=nty,maxbibnames=99]{biblatex}
\usepackage{algorithmicx,algpseudocode}
\usepackage{algorithm}

\author{Roland Wagner, Daniela Saxenhuber, Ronny Ramlau, Simon Hubmer}
\title{Direction dependent Point spread function reconstruction for Multi-Conjugate Adaptive Optics on Giant Segmented Mirror Telescopes}

\begin{document}

\maketitle\section{Introduction}
As for any optical system, the observed image $I_o$ of a ground based astronomical telescope can be described as a convolution of the true image $I$ with the so-called Point Spread Function (PSF), i.e.,
\begin{align}\label{eq:conv}
I_o = I\ast PSF.
\end{align}

For an ideal telescope, i.e., without additional aberrations from the atmosphere or imperfections of the instrument, the PSF only depends on the geometry of the telescope and is given by
\begin{equation}
  PSF_{opt}= \mathcal{F}\left(\chi_{\mathcal{P}}\right) ,
\end{equation}
where $\chi_{_{\mathcal{P}}}$ denotes the characteristic function of the telescope aperture $\mathcal{P}$ and $\mathcal{F}$ is the Fourier transform. In the following, we describe how the $PSF$ is altered from this perfect $PSF_{opt}$ in ground-based telescopes using AO systems.
Turbulence in the atmosphere above ground-based telescopes alter their PSF significantly, leading to blurred scientific images. Although modern ground-based telescopes reduce the effect of the turbulent atmosphere by Adaptive Optics (AO) systems, residual turbulence remains uncorrected, i.e., the PSF of the telescope is still degraded. Thus the goal of PSF reconstruction methods is to obtain the actual PSF $PSF_{\phi}$  from data acquired by the wavefront sensors (WFS) and the commands applied to the deformable mirror(s) (DM) in a post-processing step. The PSF after AO correction at time $t$ is given by
\begin{equation}
PSF_{\phi} = \mathcal{F}\left( \chi_{\mathcal{P}}e^{i\phi(\cdot,t)}\right),
\end{equation}
where $\phi(x,t)$ denotes the residual incoming phase.

In general, the turbulence distribution in the atmospheric volume above a ground based telescope is spatially dependent. This causes a variation in the PSF as light beams coming from different celestial objects cut through different parts of this 3D volume, %different lines of sight cut through different parts of this 3D volume, 
unless all turbulence is located very close to the ground. The PSF is thus directional dependent and changes to
\begin{equation}
PSF_{\phi,\alpha} = \mathcal{F}\left( \chi_{\mathcal{P}}e^{i\phi_\alpha(\cdot,t)}\right),
\end{equation}
where $\phi_\alpha(x,t)$ is the residual incoming phase in direction $\alpha$. This model of the PSF can now be used in \eqref{eq:conv}. Note that this models the PSF at a certain time. However, the PSF related to an observation is always an average over an exposure. We will address this in Section~\ref{sect:psfr}. We assume that the PSF is constant on so-called isoplanatic patches, i.e., the observed image can be split into small areas (patches) on which the PSF is constant. It was shown (e.g., Ref.~\cite{Br06PSF}) that such patches are only a few arcseconds in diameter. Our goal is to present a method for reconstructing a PSF for each of those isoplanatic patches.

The purpose of this paper is the development of an algorithm for the reconstruction of an exposure average of the directional dependent PSF $PSF_{\phi,\alpha}$ based on a tomography of the atmosphere in Multi-Conjugate Adaptive Optics (MCAO). Specifically, we are interested in directions which do not coincide with guide star directions. Knowledge of the PSF in different directions is required, as the field of view of several instruments for the upcoming generation of Extremely Large Telescopes will cover an area of around 1', since the PSF varies every few arcseconds. Therefore, the image quality of MCAO systems is degraded (and maybe even dominated) by the so-called tomographic and generalized fitting errors. The tomographic error originates from the atmospheric tomography being performed from a small number of guide star (GS) measurements and using a layered model for the atmosphere as opposed to the full 3D volume (see, e.g., Ref.~\cite{SaRa15}). Therefore, the true atmospheric volume cannot be reconstructed perfectly. The generalized fitting error adds to this as the reconstructed atmospheric turbulence then needs to be projected onto one to three DMs. Additionally, all AO corrected images suffer from a time delay due to the WFS integration time and the time required for the adjustment of the DM(s), which results in a larger residual wavefront. 

A possible use of the reconstructed PSF is as a quality measure for the science images. This is of particular interest when choosing the best images of an observation series when several science images are stacked together before being analyzed further. Additionally, one can use the reconstructed PSF to post-process the science images, e.g., in a deconvolution algorithm, as shown in Refs.~\cite{DyRaReSoWa21,RaSoHu21}. \\

In the past, several methods have been proposed for PSF reconstruction (for an overview cf., Ref.~\cite{WaRaLAM19}). % In contrast, our method uses a tomographic reconstruction of the atmosphere to obtain direction dependent PSFs and does not need such a precomputed filter.\\
Most methods for PSF reconstruction purely from AO telemetry data start from the work in Ref.~\cite{VeRiMaRo97PSF}, where a method using pre-computed so-called $U_{i,j}$-functions was presented.
In Ref.~\cite{GeClFuRo06PSF}, a simplification of this original work in Ref.~\cite{VeRiMaRo97PSF} was proposed: for each observation the eigen decomposition of the representation in $U_{i,j}$-functions is computed on the fly. This results in a huge reduction of the computational costs and the required memory, but needs an eigenvalue decomposition of the parallel phase covariance matrix on the fly to replace the pre-computed functions. 
The method proposed in  Ref.~\cite{JoVeCo2006PSF} tries to model the phase power spectrum analytically, and in Ref.~\cite{ExGrRo13PSF} a maximum likelihood approach for the used covariance matrices is taken. 
%In Ref.~\cite{NaJeCh10} only open loop data is taken into account and a fine resolution wavefront is created by combining measurements from different timesteps. 
These algorithms all suffer from anisoplanatism, which is accounted for in Refs.~\cite{Br06PSF,AuRoSc07PSF,BeCoMi18PSF}. The only existing method for PSF reconstruction for MCAO systems\cite{GiCoVeWaEl12PSF,GiWaBo18} relies on WFS data and use global basis functions for the reconstruction. In this methods, the structure function of the incoming wavefront is extrapolated from the structure function calculated from wavefront sensor data. The filter for this extrapolation is obtained from simulations as the ratio between the structure functions in guide star direction and in the direction of interest. 
Please note that all the mentioned methods - with the exception of Refs.~\cite{GiCoVeWaEl12PSF,GiWaBo18} - were designed for Single Conjugate Adaptive Optics systems (SCAO) only. Some of the algorithms were already successfully tested on sky on various telescopes.\cite{JoNe12PSF,JoChWiTo10PSF,Fl08PSF,ClKaGe06PSF,BeFeNeFu20PSF,BeMaFu20PSF} Additionally, methods combining focal and pupil plane measurements to derive PSFs were developed in recent years\cite{BeCoRa19PSF}. As pointed out in Ref.~\cite{DaKa12PSF}, before the algorithms for reconstructing the PSF existed, the PSF of a reference star was measured and then used for deconvolution algorithms. \\

Complementary to the existing methods, our approach will allow to reconstruct a PSF for any direction in the field of view from the AO telemetry data by using a tomographic approach where no extrapolations are needed. Furthermore, our method, which is based on Ref.~\cite{WaHoRa18}, neither needs pre-computed functions nor an eigenvalue decomposition of a matrix on the fly in contrast to the methods from Ref.~\cite{VeRiMaRo97PSF} and Ref.~\cite{GeClFuRo06PSF}, respectively. We briefly describe it in the following paragraph.

Our PSF reconstruction method will be based on WFS data which is acquired at a frequency of around 500 Hz. %, which results in a large amount of data gathered for a single observation. 
Even for the upcoming generation of instruments, PSF reconstruction will be performed {\it after} the observation. As the image exposure time ranges from a few seconds to several minutes, this may result in a large amount of data to be processed. Therefore, an algorithm for PSF reconstruction has to be computationally efficient and should be adapted to the available hardware, in particular with respect to memory consumption.\\

The proposed algorithm for MCAO PSF reconstruction, whose basic idea was first described in a PhD thesis\cite{Wa17}, overcomes the difficulties arising from the spatial dependence of the PSF in two steps. First, a tomographic reconstruction of the atmosphere to compute the direction dependent residual wavefront is performed. For this, we will use the so-called 3-step method from Ref.~\cite{SaRa15}. Second, this residual wavefront is then used as an input to the classical PSF reconstruction algorithm from Ref.~\cite{VeRiMaRo97PSF}, modified to make use of a 4D structure function instead of a 2D version and adapted to bilinear basis functions based on the work presented in Ref.~\cite{WaHoRa18}. Note that the method from Ref.~\cite{WaHoRa18} neither needs pre-computed functions nor an eigenvalue decomposition of a matrix on the fly in contrast to the methods from Ref.~\cite{VeRiMaRo97PSF} and Ref.~\cite{GeClFuRo06PSF}, respectively.

As a consequence of the discrete resolution of any WFS and the limited correction by the DM(s), high spatial frequencies of the incoming phase, which are forming the wings of the PSF outside the so-called control radius, have to be {\it simulated} using a statistical model of the atmosphere. Such a simulation and the calculation of the respective structure function has to be performed on a finer grid than the WFS if bilinear splines are used as basis functions. Furthermore, a model of the noise influencing the measurements has to be included as well.\\

The computation of the directional dependent wavefront is based on the measured wavefront in guide star direction as well as on the current mirror shapes. Therefore, we need to use a reliable inversion method for atmospheric tomography in order to reconstruct the atmospheric layers as well as projections through the atmosphere in order to obtain estimates of the wavefronts for each direction of interest for the  PSF reconstruction. The proposed algorithm will be able to provide a number $N_{PSF}$ of reconstructed PSFs in directions $\beta_i$, $i = 1,\dots, N_{PSF}$, which may be different from the guide star directions $\alpha_g$, $g = 1, \dots, G$. Using this tomographic approach will help to account for the correlation between bandwidth and anisoplanatic errors in the PSF, which has been identified in a careful analysis of the error budget for an SCAO system in Ref.~\cite{FeGeRoGr18PSF}.\\

The remainder of this paper is structured as follows: In Section~\ref{sect:psfr}, we recall the basics for PSF reconstruction from AO telemetry data. In Section~\ref{sect:rec-atm}, we introduce the tomographic reconstruction algorithm from Ref.~\cite{SaRa15}. Our new method for PSF reconstruction in MCAO systems is presented in Section~\ref{sect:method}. Finally, in Section~\ref{sect:numerics}, we demonstrate the feasibility of our method using data from the European Southern Observatory's (ESO) end-to-end simulation tool OCTOPUS\cite{OCTOPUS}.

\section{Background on PSF reconstruction from AO telemetry data}\label{sect:psfr}
To understand the principles of PSF reconstruction using only AO telemetry data, which is also called pure PSF reconstruction, let us start from the instantaneous optical transfer function (OTF). In the near field approximation for a monochromatic image, the OTF at wavelength $\lambda$ and time $t$, denoted by $B$, is given in Ref.~\cite{VeRiMaRo97PSF} as
\begin{equation}
B({\boldsymbol \rho}/\lambda,t) = \frac{1}{S} \iint_\mathcal{P} P({\bf x}) P({\bf x}+{\boldsymbol \rho})\, \text{exp}\left(i \phi({\bf x},t)-i \phi({\bf x}+{\boldsymbol \rho},t)\right)d{\bf x},
\end{equation}
where $S$ is the area of the telescope aperture, $P$ is the pupil function and $\phi$ is the residual phase after the AO correction. In case of a perfect optical system, it holds that $P = \chi_\mathcal{P}$. The normalization ensures that the PSF has unit energy. Note that the dependence on the wavelength $\lambda$ is implicit through the relation between phase $\phi$ and wavefront $\varphi$, given as $\phi(x,t) = \frac{2\pi}{\lambda} \varphi(x,t)$.  Averaging the instantaneous OTF $B({\boldsymbol \rho}/\lambda,t)$ over the integration time interval gives the long exposure OTF $B({\boldsymbol \rho}/\lambda) $.\\

Furthermore, we assume that the corrected phase $\phi$ at any position on the pupil has a Gaussian statistics and the integration time is long enough such that the statistical average can be substituted by the temporal average. This leads to the following expression of the long exposure OTF:
\begin{equation}\label{OTF}
B({\boldsymbol \rho}/\lambda) = \langle B({\boldsymbol \rho}/\lambda,t) \rangle_t =  \frac{1}{S} \iint_\mathcal{P} P({\bf x}) P({\bf x}+{\boldsymbol \rho})\, \text{exp}\left(-\frac{1}{2}D_{\phi}({\bf x},{\boldsymbol \rho})\right)  d{\bf x},
\end{equation}
where the structure function of the residual incoming phase is defined as
\begin{align}\label{eq:sf}
D_\phi({\bf x},{\boldsymbol \rho}) = \langle |\phi({\bf x},t) - \phi ({\bf x} + {\boldsymbol \rho},t)|^2\rangle_t,
\end{align}
with $\langle \cdot \rangle_t$ denoting the temporal average of a function.
With this, one can obtain the long exposure PSF by applying the Fourier transform to the long exposure OTF, i.e.,
\begin{equation*}
PSF({\bf u }) = \mathcal{F}(B({\boldsymbol \rho}/\lambda)).
\end{equation*}

\subsection{Modern PSF reconstruction for SCAO}
In order to compute the PSF in an SCAO system, we start from Eq.~\eqref{OTF}. The exact calculation of $B({\boldsymbol \rho}/\lambda)$ requires averaging four dimensional functions, which seemed computationally too demanding when the first algorithm for PSF reconstruction from AO data was developed in Ref.~\cite{VeRiMaRo97PSF}. Thus, it was proposed to interchange spatial average and the exponential function in Eq.~\eqref{OTF} to overcome this problem. Today this simplification can be dropped as the computation of the structure function from AO telemetry data is possible in reasonable time even on a laptop as shown, e.g., in Ref.~\cite{GiCoVeWaEl12PSF}.\\

As a starting point of the approach introduced in Ref.~\cite{WaHoRa18}, $\phi$ is split into a part seen by the WFS (and thus corrected by the DM in the subsequent time step of the AO system), called $\phi_\|$, also denoted as lower order part, and a part orthogonal to the modes sensed by the WFS, called $\phi_\perp$, denoted as higher order part. Clearly, $\phi = \phi_\| + \phi_\perp$ and thus from Eq.~\eqref{eq:sf} we get:
\begin{equation}\label{eq:sf_split}
D_\phi ({\bf x},{\boldsymbol \rho})= D_{\phi_\|}({\bf x},{\boldsymbol \rho}) + D_{\phi_\perp} ({\bf x},{\boldsymbol \rho})+ 2 \langle [\phi_\|({\bf x},t) -\phi_\| ({\bf x} + {\boldsymbol \rho},t)][\phi_\perp({\bf x},t) - \phi_\perp ({\bf x} + {\boldsymbol \rho},t) ] \rangle_t.
\end{equation}
Note that in Ref.~\cite{VeRiMaRo97PSF} this splitting was done after interchanging the spatial average and the exponential in Eq.~\eqref{OTF}.\\
Furthermore, note that the last term of Eq.~\eqref{eq:sf_split} is not an inner product, but a temporal average of the point-wise product of two orthogonal functions. Therefore, this term does not vanish in general. However, we follow Ref.~\cite{VeRiMaRo97PSF} and neglect it. 
Thus, the OTF can be approximated by
\begin{equation*}
B({\boldsymbol \rho}/\lambda) =  \frac{1}{S} \iint_\mathcal{P} P({\bf x}) P({\bf x}+{\boldsymbol \rho})\, \text{exp}\left(-\frac{1}{2}D_{\phi_\|}({\bf x},{\boldsymbol \rho})\right) \text{exp}\left(-\frac{1}{2}D_{\phi_\perp}({\bf x},{\boldsymbol \rho}) \right)  d{\bf x}.
\end{equation*}

The orthogonal part of the residual phase cannot be measured from the actual on sky data, but only simulated offline by using sophisticated atmospheric models. Therefore, a calculation of the orthogonal part structure function $D_{\phi_\perp}({\bf x},{\boldsymbol \rho})$ gives no physically meaningful contribution and we follow the suggestion of Ref.~\cite{VeRiMaRo97PSF} to replace it by its mean over the variable ${\bf x}$, i.e.,
\[
D_{\phi_\perp}({\bf x},{\boldsymbol \rho}) \approx \bar{D}_{\phi_\perp}({\boldsymbol \rho}) = \frac{\iint_{\mathcal{P}} P({\bf x}) P({\bf x}+{\boldsymbol \rho}) D_{\phi_\perp}({\bf x},{\boldsymbol \rho}) d{\bf x}}{\iint_{\mathcal{P}} P({\bf x}) P({\bf x}+{\boldsymbol \rho}) d{\bf x}},
\]
so that Eq.~\eqref{OTF} simplifies to
\begin{equation}\label{eq:otf_prod}
B({\boldsymbol \rho}/\lambda)  = \frac{1}{S} \iint_\mathcal{P} P({\bf x}) P({\bf x}+{\boldsymbol \rho})\, \text{exp}\left(-\frac{1}{2}D_{\phi_\|}({\bf x},{\boldsymbol \rho})\right) d{\bf x} \cdot \text{exp}\left(-\frac{1}{2}\bar{D}_{\phi_\perp}( {\boldsymbol \rho} ) \right).
\end{equation}
As Eq.~\eqref{eq:otf_prod} is a product of two independent terms, we define
\begin{align}
& B_{\|+tel}({\boldsymbol \rho}/\lambda) := \frac{1}{S} \iint_\mathcal{P} P({\bf x}) P({\bf x}+{\boldsymbol \rho})\, \text{exp}\left(-\frac{1}{2}D_{\phi_\|}({\bf x},{\boldsymbol \rho})\right) d{\bf x}, \label{eq:otf_partel} \\
& B_{\perp}({\boldsymbol \rho}/\lambda) := \text{exp}\left(-\frac{1}{2}\bar{D}_{\phi_\perp}( {\boldsymbol \rho} ) \right). \label{eq:otf_perp}
\end{align}
The first term $B_{\|+tel}({\boldsymbol \rho}/\lambda)$ has to be calculated from closed loop AO measurements either on the fly or in a post-processing step and the second term $B_{\perp}({\boldsymbol \rho}/\lambda)$ can only be estimated from simulation, as no information on $\phi_\perp$ is available in on-sky telemetry data. Note that when using the original method from Ref.~\cite{VeRiMaRo97PSF} also the structure function of $\phi_\|$ is averaged over ${\bf x}$, which results in three independent components: the OTF of the telescope in the absence of turbulence, the contribution of the mirror component and the contribution of the higher order phase. In contrast, in our approach the first two components are combined into $B_{\|+tel}({\boldsymbol \rho}/\lambda)$. \\

For the computation of $ \bar{D}_{\phi_\perp}({\boldsymbol \rho})$, a Monte Carlo method is proposed e.g., in Ref.~\cite{VeRiMaRo97PSF}, where the high order component of randomly generated phase screens with Kolmogorov or Van Karman statistics are extracted and then used to estimate the structure function (see, e.g., Ref.~\cite{WaHoRa18}).

\subsection{Choice of the basis functions}
To be able to compute the parallel part structure function $D_{\phi_\|}(x, {\boldsymbol \rho})$ in an effcient way, V\'{e}ran\cite{VeRiMaRo97PSF} introduced functions $U_{ij}({\boldsymbol \rho})$ which depend on the modes of the DM and can be precomputed numerically. For the remaining calculation only the time averaged covariances of these modes are needed. The original algorithm was proposed for Zernike polynomials and lead to good results, but is time and memory consuming. \\

As Zernike polynomials have a global support, the computation of the $U_{ij}$-functions requires the assembly of full matrices, which leads to increased memory consumption and requires more computational power. Due to the high degrees of freedom for the future Giant Segmented Mirror Telescopes (GSMT) this is not feasible, as it would require to use more than the first 4000 Zernike polynomials. In particular, if one considers the use of (almost) linear influence functions for future DMs, one could think of using, e.g., bilinear splines as basis functions for the $U_{ij}$-algorithm as proposed in Refs.~\cite{WaHoRa18,Ho14}. This change leads to a sparse representation of the required matrices for the mirror part. However, if these functions are defined on the actuator positions only, higher order terms cannot be represented. Therefore, we define the bilinear functions on a finer grid with a spacing of, e.g., $1/4$ of the DM actuator spacing (see Ref.~\cite{WaHoRa18}).
%To overcome this problem, we use a finer resolution of the wavefronts for the estimation of the higher order parts (see Ref.~\cite{WaHoRa18}). 
As a drawback, this results in higher memory consumption and slower computations. Since the higher order components $\phi_\perp$ are precomputed and appropriately scaled to current observation conditions, this is not a crucial issue with modern computers. Note that any choice of basis functions will lead to similar needs in terms of degrees of freedom.\\

After recalling these preliminaries, we proceed with a short review of the existing PSF reconstruction algorithms for MCAO.

\subsection{Extending PSF reconstruction to MCAO}
To our knowledge, so far only Gilles et al.\cite{GiCoVeWaEl12PSF,GiWaBo18} have developed an algorithm for PSF reconstruction in an MCAO system. Their approach is an extension of PSF reconstruction for an SCAO system using Laser Guide Stars (LGS). The structure function for the LGS direction is decoupled into a tip-tilt structure function and a tip-tilt-removed structure function, due to the fact that a WFS attached to an LGS is insensitive to tip-tilt. The extension in Refs.~\cite{GiCoVeWaEl12PSF,GiWaBo18} to other directions than the LGS ones is done by a multiplicative filter. Such a filter is obtained as the balanced or unbalanced ratio between structure functions obtained for the LGS direction and the science direction from simulations. 
Note that this approach is not using a tomographic reconstruction of the atmosphere, but only extrapolating the structure function from guide star directions into other directions.
%We want to highlight that this approach does not use a tomographic reconstruction of the atmosphere, which allows an estimation of the incoming wavefront in the direction of interest without using any simulations. It is only based on reconstructed structure function in guide star direction accompanied with a direction dependent filter from simulated data. 
Therefore, the existing method is in clear contrast to our proposed algorithm.

\section{Reconstruction of the turbulent atmosphere}\label{sect:rec-atm}
An integral part of our PSF reconstruction method for MCAO is the usage of atmospheric tomography, which we describe in the following paragraphs.\\

Wide field of view AO systems rely on a good atmospheric reconstruction which is then used to obtain optimal mirror shapes for the AO correction. In MCAO systems, several DMs conjugated to different altitudes are used to yield a good correction over the whole field of view. As input, Shack-Hartmann wavefront sensor (SH WFS) data from $G$ guide stars (natural and/or laser) are used. In this section, we introduce our method of choice for reconstructing the turbulent atmosphere and describe how it is used to derive a phase as input to the PSF reconstruction algorithm.\\

To obtain a tomographic reconstruction of the atmosphere, several methods exist. In Ref.~\cite{Fusco}, WFS data are mapped directly on the DMs. Most methods split the problem of determining the DM shapes from WFS data into two separate and independent subproblems. Members of this family of methods are the Fractal Iterative Method (FrIM) \cite{ThiTa10,TaBeTaLeThClMa11} and the Finite Element Wavelet Hybrid Algorithm (FEWHA) \cite{YuHeRa13b,Yu14,HeYu13,YuHeRa13}. They all have in common that they first reconstruct the  atmospheric turbulence directly from the WFS data, before deriving the shapes of the DMs from this reconstruction. A new generation of tomographic methods splits the MCAO problem into three steps. This results in  separate steps for wavefront reconstruction from WFS data and the tomography to obtain atmospheric layers and are therefore called three-step methods. For a comparison between classical two-step methods and the 3-step approach, see, e.g., Ref.~\cite{RaSaYu14}.

\subsection{The 3-step approach}\label{sect:3step}
For the tomographic reconstruction we use the 3-step approach with the Gradient-based method developed in Refs.~\cite{SaRa15,RaRo12}, which we briefly introduce in this section.\\

In the first step, the incoming wavefronts ${\boldsymbol \varphi} = (\varphi_1, \dots ,\varphi_G)$ are reconstructed from SH WFS data using the Cumulative Reconstructor with Domain decomposition (CuReD) \cite{Ros11,Ros12}, which is a very efficient, direct wavefront reconstruction method. Alternatively one could also use the Cumulative Reconstructor on a Finite Element basis with Domain decomposition (FinECuReD) \cite{Neub12b,WaNeRa17}, or the hierarchical wavefront reconstruction algorithm (HWR) \cite{BhBiBaMyDi13}. \\

In the second step, $L$ turbulent layers of the atmosphere ${\boldsymbol \Phi} = (\Phi^{(1)}, \dots , \Phi^{(L)})$ are reconstructed from incoming wavefronts. This is the so-called atmospheric tomography problem (starting from wavefronts), where we need to solve a system of equations
\begin{align}\label{eq:opeq}
{\boldsymbol A} {\boldsymbol \Phi} = {\boldsymbol \varphi}\,,%\left(\begin{array}{c}\A{1}\\ \vdots\\ \A{G} \end{array} \right) {\boldsymbol \Phi} =   \left(\begin{array}{c}\vphi{1}\\ \vdots\\ \vphi{G}\end{array} \right), 
\end{align}
where ${\boldsymbol A} = (A_1,\dots ,A_G)$  describes projection operators in $G$ guide star directions. The projection $A_g$ is defined as 
\begin{equation}
\label{eq:Ag}
A_g{\boldsymbol \Phi} = \sum_{l=1}^L \Phi^{(l)}(c_lr + \alpha_g h_l),
\end{equation}  
with spatial coordinates on the aperture $r = (x,y)$, the direction $\alpha_g $ in arcmin, $h_l$ the height of layer $l$ and the LGS scaling factor $c_l = 1-h_l/h_{LGS}$ for LGS at height $h_{LGS}$ and $c_l=1$ for NGS. Please note that the projection operators $A_g$ coincide with the standard Radon transform on a layered medium. \\

 Equation \eqref{eq:opeq} is solved by the minimization of the least squares functional, i.e.,
\begin{align*}
{\boldsymbol \Phi}^{rec} = \mbox{argmin}_{\boldsymbol \Phi} ||{\boldsymbol A}{\boldsymbol \Phi} - {\boldsymbol \varphi}||^2\,.
\end{align*}
As described in Ref.~\cite{SaRa15}, a Gradient-based iteration 
\begin{equation}
{\boldsymbol \Phi}_{j+1} = {\boldsymbol \Phi}_j + \tau_j{\boldsymbol A}^\ast ({\boldsymbol \varphi} - {\boldsymbol A}{\boldsymbol \Phi}_j)\,,
\end{equation} 
where ${\boldsymbol A}^\ast$ denotes the adjoint of the projection ${\boldsymbol A}$ and $\tau_j$ is a stepsize, is used for minimization. \\

The third step is to compute the shapes of the $M$ DMs from the reconstructed atmosphere. In the Gradient-based method, this is done by minimizing the following functional:
\begin{align*}
\mbox{argmin}_{{\boldsymbol \Phi}_{DM}} \|\tilde{{\boldsymbol A}}{\boldsymbol \Phi}_{DM} - {\boldsymbol A}{\boldsymbol \Phi^{rec}} \|^2,
\end{align*}
with $\tilde{{\boldsymbol A}} = (\tilde{{\boldsymbol A}}_{1}, \dots, \tilde{{\boldsymbol A}}_{D})$ the concatenated tomography operators for the optimization directions $\tilde{\alpha}_d$, $d = 1, \dots, D$ and the DM heights $\tilde{h}_m$, $m=1,\dots,M$, as introduced in Refs.~\cite{RaRo12,RoRa13,RafRaYu16}, i.e.,
\begin{equation}
\label{eq:Atildealpha}
\tilde{A}_{d}{\boldsymbol \Phi} = \sum_{m=1}^M \Phi^{(m)}(c_mr + \tilde{\alpha}_d \tilde{h}_m).
\end{equation}

 \subsection{Pseudo projection step for PSF reconstruction}\label{sect:pseudo-proj}
For our PSF reconstruction method, we need as input the residual wavefronts $\varphi_{\|,\beta_i}$ in directions $\beta_i$, $i = 1,\dots,N_{PSF}$, which describe the residual blur after AO correction in the respective direction. However, due to noise and aliasing effects, we cannot obtain $\varphi_{\|,\beta_i}$, but only an approximation of the wavefront, denoted as $\varphi_{rec,\beta_i}$.
%After reconstruction of the turbulent atmosphere, we want to obtain the wavefronts $\varphi_{\|,\beta_i}$ in directions $\beta_i$, $i = 1,\dots,N_{PSF}$. 
As these wavefronts are not related to celestial WFS measurements, we call them \emph{pseudo wavefronts}. The idea is to obtain $\varphi_{rec,\beta_i}$ from the reconstructed atmospheric layers ${\boldsymbol \Phi}^{rec}$ using the projection operator. Note that in all AO systems one will only obtain an approximation for the residual wavefront due to noise and aliasing errors. \\

Since the definition of the projection operator in Eq.~\eqref{eq:Ag} is not restricted to guide star directions, we can also use it in the directions $\beta_i$ as
\begin{equation}
\label{eq:Abeta}
A_i{\boldsymbol \Phi} = \sum_{l=1}^L \Phi^{(l)}(c_lr + \beta_i h_l).
\end{equation}  
This enables us to obtain the reconstructed pseudo wavefronts $\varphi_{rec,\beta_i}^{ol}$ from the reconstructed atmospheric layers ${\boldsymbol \Phi}^{rec}$ as
%Having the reconstructed atmospheric layers ${\boldsymbol \Phi}^{rec}$ at hand, it is straight forward to obtain the pseudo wavefronts $\varphi_{\|,\beta_i}$, by using the projection operator, defined in Eq.~\eqref{eq:Ag}, with $\beta_i$ instead of $\alpha_g$. The resulting pseudo wavefronts 
\begin{align}\label{eq:mcao-pseudo-proj}
 \varphi_{rec,\beta_i}^{ol} = A_i {\boldsymbol \Phi}^{rec}, \qquad i= 1,\dots, N_{PSF},
\end{align} 
where the $\varphi_{rec,\beta_i}^{ol}$ are in open loop. Hence, we need to remove the current DM shapes by
\begin{align}
\varphi_{rec,\beta_i} = \varphi_{rec,\beta_i}^{ol}-\tilde{{\boldsymbol A}}{\boldsymbol \Phi}_{DM},
\end{align}
before further using $\varphi_{rec,\beta_i}$ in the calculation of the OTF, which relies on residual wavefronts only.%\\

%If the implemented reconstruction method for MCAO on a telescope relies on a reconstruction of the atmospheric layers, e.g., Refs.~\cite{RaRo12,RoRa13,YuHeRa13b,RaSaYu14}, the projection step for the pseudo wavefronts can be directly implemented there and the pseudo wavefronts can be saved for the further processing after sending the current commands to the DMs. For GSMTs, like the ELT of ESO currently under construction, the aim is to get a reconstructed PSF in several different directions using the pseudo projection step.

\section{PSF reconstruction in an MCAO system}\label{sect:method}
In this section, we present our algorithm for PSF reconstruction in an MCAO system. \\

Regardless of the employed AO system, the considerations of Section~\ref{sect:psfr} are always valid in case of aberrations. In particular, Eq.~\eqref{eq:otf_prod} remains the starting point for PSF reconstruction also in an MCAO system.\\

However, MCAO systems do not necessarily have an on-axis guide star. This is, e.g., the case for MAORY\cite{MAORY20}, the MCAO system under construction for ESO's Extremely Large Telescope (ELT). Therefore, no on-axis WFS measurements might be available, leading to a lack of information for PSF reconstruction in the center direction. Additionally, as MCAO covers a wide FoV, one might also aim for knowledge of the PSF in several positions in the FoV to account for the spatial variability.\\

Let us denote by $N_{PSF}$ the number of PSFs which we want to reconstruct, and by $\beta_i$, $i = 1,\dots, N_{PSF}$ the directions corresponding to the PSF in the FoV. In order to use Eq.~\eqref{eq:otf_prod}, we need the phase $\phi_{\|,\beta_i}$, or equivalently the wavefront $\varphi_{\|,\beta_i}$, in each direction. As mentioned above, our approach is to use the reconstructed atmosphere from the real time control algorithm for the DM control for this purpose.\\

Using Eq.~\eqref{eq:otf_prod} for the directions $\beta_i$ under consideration, which leads to a different OTF for each direction $\beta_i$, we obtain 
\begin{align}
& B_{\beta_i}({\boldsymbol \rho}/\lambda) = \frac{1}{S}  B_\perp({\boldsymbol \rho}/\lambda)   \cdot  B_{\|+tel,\beta_i}({\boldsymbol \rho}/\lambda) , \label{eq:psfr-mcao} \\
&  B_\perp({\boldsymbol \rho}/\lambda)  :=  \text{exp}\left(-\frac{1}{2}\bar{D}_{\phi_\perp}( {\boldsymbol \rho} ) \right), \label{eq:mcao-otf_perp} \\
&  B_{\|+tel,\beta_i}({\boldsymbol \rho}/\lambda) := \int_\mathcal{P} P({\bf x}) P({\bf x}+{\boldsymbol \rho})\, \text{exp}\left(-\frac{1}{2}D_{\phi_{\|,\beta_i}}({\bf x},{\boldsymbol \rho})\right) d{\bf x} \label{eq:mcao-otf_par}.
\end{align}
The PSF is then obtained as $PSF({\boldsymbol u}) = \mathcal{F}(B({\boldsymbol \rho}/\lambda)$.\\

As for an SCAO system, one still has to obtain good estimates for $ \bar{D}_{\phi_\perp}( {\boldsymbol \rho} )$ in Eq.~\eqref{eq:mcao-otf_perp}, and $D_{\phi_{\|,\beta_i}}({\bf x},{\boldsymbol \rho})$ in Eq.~\eqref{eq:mcao-otf_par}. Again, the first term can only be estimated from simulation. The procedure for computing $D_{\phi_{\|,\beta_i}}({\bf x},{\boldsymbol \rho})$ now has to use the tomographic approach from Section~\ref{sect:3step} combined with Section~\ref{sect:pseudo-proj}, as for $\phi_{\|,\beta_i}$ no WFS measurements need to be available.

\subsection{Calculating the structure functions in an MCAO system}
As for the SCAO case, the estimated structure functions have to be combined with the simulated part for the higher order terms in Eq.~\eqref{eq:psfr-mcao} after the exposure time.  Even though the PSF is spatially varying, the higher order component can only be simulated as a spatial average related to the statistical distribution of the uncorrected frequencies of the residual phase under the current atmospheric conditions. Therefore, it is a viable way to use the same higher order structure function for each direction. This helps to keep our method reasonable also in terms of computational power and required memory space.

As in Refs.~\cite{WaHoRa18,VeRiMaRo97PSF}, we decompose the reconstructed incoming wavefront $\phi_{rec,\beta_i}$ into $\phi_{rec,\beta_i} = \phi_{\|,\beta_i} + \phi_{n,\beta_i} + \phi_{r,\beta_i}$. Note that $\phi_{rec,\beta_i}$ is only an estimate for $\phi_{\|,\beta_i}$ and $\phi_{n,\beta_i}$ and $\phi_{r,\beta_i}$ account for different errors. In particular, $\phi_{n,\beta_i}$ is the WFS noise propagated into the pseudo-wavefront $\phi_{rec,\beta_i}$ and $\phi_{r,\beta_i}$ is the higher order component giving a non-zero measurement and being propagated into DM commands. The latter is known as aliasing, i.e., $\phi_{r,\beta_i} = R_i \Gamma \phi_\perp$, with $\Gamma$ the SH-WFS operator and $R_i$ the concatenation of the tomography and projection operator for direction $\beta_i$, which coincides with the AO control algorithm in guide star directions.

Additionally, we need to model the structure functions for noise $\phi_n$ and aliasing $\phi_r$ separately, as they come from different sources. This results in a splitting of the structure function for the parallel part, using the same ideas as in Ref.~\cite{VeRiMaRo97PSF} and assuming that noise and aliasing are independent and stationary, i.e.,
\begin{align}\label{eq:Dphipar}
D_{\phi_{\|,\beta_i}} ({\bf x},{\boldsymbol \rho}) \approx D_{\phi_{rec,\beta_i}}({\bf x},{\boldsymbol \rho})- \bar{D}_{\phi_{n,\beta_i}}({\boldsymbol \rho}) + \bar{D}_{\phi_{r,\beta_i}}({\boldsymbol \rho}).
\end{align}

Let us describe how the three terms in Eq.~\eqref{eq:Dphipar} can be computed. First, the structure function $D_{\phi_{rec,\beta_i}}$ relies on the computed pseudo wavefronts $\varphi_{rec,\beta_i}$, via the corresponding phase $\phi_{rec,\beta_i} = \frac{2\pi}{\lambda}\varphi_{rec,\beta_i}$, in the desired directions and can be directly computed from them.

As in Ref.~\cite{WaHoRa18}, the structure functions for noise and aliasing, $ \bar{D}_{\phi_{n,\beta_i}}$ and $ \bar{D}_{\phi_{r,\beta_i}}$, are spatially averaged from realizations of $\phi_{n,\beta_i}$ and $\phi_{r,\beta_i}$ from Monte Carlo simulations, respectively. We assume a Gaussian white noise covariance matrix on the wavefront sensor, i.e., $C_{WFS} = \frac{1}{n_{photons}}I$. To obtain $\phi_{r,\beta_i}$, we simulate $\phi_\perp$, compute the respective SH measurements $\Gamma\phi_\perp$ and use the same algorithm as for PSF reconstruction in an SCAO system. In particular, this means that we need to propagate the aliasing term through the full atmospheric tomography and the projection step in the desired direction $\beta_i$. Furthermore, note that the noise and aliasing structure functions are computed only once as a starting point and updates can be performed offline, so one could use available covariance matrices together with the $U_{ij}$-functions as proposed in Ref.~\cite{VeRiMaRo97PSF}. However, when using a matrix-free AO control algorithm this would mean that one needs to set up the corresponding matrix, which could easily be done by computing the response of the algorithm to only one non-zero measurement.

\subsection{Algorithm for PSF reconstruction in an MCAO system}
We summarize the algorithm for PSF reconstruction in an MCAO system in Algorithm~\ref{alg:psfr-mcao}. It consists of the following three steps: The first step, computing $B_{\|+tel,\beta_i}$, has to be performed during or after the exposure time, depending on the availability of memory and computational power. The second step, computing $B_\perp$, can be done in simulation only and therefore on a separate machine. The only required inputs are the pupil mask and the atmospheric conditions during the observation. The third step of combining the first two is post processing after the exposure time. As an additional input the directions of interest for the PSF reconstruction have to be defined. Within all these steps, we use the models for the structure functions from Ref.~\cite{WaHoRa18}. 

\begin{algorithm}[!htb]
 {\bf Input:} WFS data $s$, statistics of the noise $\eta$, Fried parameter $r_0$, directions of interest $\beta_i$  \\
 {\bf Output:} the long exposure PSF of the residual phase in direction $\beta_i$, $PSF_{\beta_i}$.\\\\
 \emph{Calculations  on AO telemetry data} \\
  {\bf For:} all telemetry frames of the exposure {\bf do} \\
\hphantom{em} get WFS data ${\bf s}$\; \\
\hphantom{em} reconstruct atmosphere ${\boldsymbol \Phi}$ as described in Section~\ref{sect:3step}\; \\
\hphantom{em} project the atmosphere onto the pupil plane in directions $\beta_i$ to obtain $\varphi_{rec,\beta_i}$ by Eq.~\eqref{eq:mcao-pseudo-proj} \; \\
 \hphantom{em}  remove the corresponding mirror correction from the projection (see Section~\ref{sect:pseudo-proj})\; \\
\hphantom{em}  calculate the 4D structure function $D_{\phi_{\| , \beta_i}}({\boldsymbol \rho}/\lambda)$ for each direction $\beta_i$ from Eq.~\eqref{eq:sf} \; \\
\hphantom{em}  calculate $B_{\|+tel,\beta_i}({\boldsymbol \rho}/\lambda)$ for each direction from Eq.~\eqref{eq:mcao-otf_par}\; \\\\
 \emph{Calculations on simulated data}\\
generate a large number of random phase screens using a von Karman or Kolmogorov statistic for the specific atmospheric conditions and extract $\phi_\perp$ by filtering out all frequencies corrected by the AO system\; \\
Compute $\bar{D}_{\phi_\perp}({\boldsymbol \rho})$ from $\phi_\perp$ using Eq.~\eqref{eq:sf} \; \\
calculate the OTF $ B_\perp({\boldsymbol \rho}/\lambda)$ from Eq.~\eqref{eq:mcao-otf_perp}\; \\\\
 \emph{Combination of the OTFs} \\
compute the OTF $B_{\beta_i} ({\boldsymbol \rho}/\lambda)$ from Eq.~\eqref{eq:psfr-mcao} for each direction $\beta_i$\; \\
obtain $PSF_{\beta_i} (u) =  \mathcal{F} (B_{\beta_i} ({\boldsymbol \rho}/\lambda))$ \;
 \caption{Point Spread Function Reconstruction for MCAO systems}
 \label{alg:psfr-mcao}
\end{algorithm}

\section{Numerical results from Octopus simulations}\label{sect:numerics}
\subsection{Simulated MCAO system}
To verify the proposed algorithm, we tested it in ESO's end-to-end simulation tool OCTOPUS \cite{OCTOPUS} for a MAORY-like ELT MCAO setting using 6 LGS WFS, with $84\times 84$ subapertures each, and 3 tip/tilt (TT) NGS. Choosing OCTOPUS as simulation tool has been natural due to our previous experience in simulating MCAO for the ELT\cite{SaRa15}. Please note that the final specifications for the MAORY system are not yet available. \\

The atmosphere used for the tests is the so-called ESO standard atmosphere with 9 layers and the seeing parameter $r_0 = 12.9~cm$ at $500~ nm$.  The simulated MCAO system, using 6 LGS and 3 TT NGS, is described in Table~\ref{tab:mcao-system}. Our simulation is based on the originally planned 42~m telescope setting for the ESO ELT. For correction of the turbulence in the atmosphere, we use three DMs, conjugated to altitudes of 0, 4000 and 12700~m. \\

\begin{table}[htbp]
\caption{Description of the simulated MCAO system}	
\label{tab:mcao-system}
\begin{center}
\begin{tabular}{l|c|c}
telescope diameter &\multicolumn{2}{c}{42~m } \\
\hline
central obstruction & \multicolumn{2}{c}{11.76~m} \\
\hline
Sodium (Na) layer altitude & \multicolumn{2}{c}{90000~m} \\
\hline
Sodium (Na) layer FWHM & \multicolumn{2}{c}{11400~m} \\
\hline
WFS integration time & \multicolumn{2}{c}{2~ms} \\
\hline
3 DM at heights 0~m, 4000~m and 12700~m & \multicolumn{2}{c}{closed loop} \\
\hline
 DM actuator spacing & \multicolumn{2}{c}{0.5~m, 1~m, 1~m} \\
\hline
Guidestars & LGS & NGS \\
\hline
Shack-Hartmann WFS & 6 & 3  \\
\hline
subapertures per WFS & $84\times84$ & $1\times 1$\\
\hline
%WFS subaperture FoV & 2.4~arcsec & 0.64~arcsec \\
%\hline
%pixels per subaperture & 16 & 8 \\
%\hline
WFS wavelength $\lambda$ & $0.589~\mu m $  & $1.65~\mu m $ \\
\hline
spot elongation & OFF  & -- \\
\hline
%spot FWHM & 1.1''  & -- \\
%\hline
detector read out noise &3e/pixel/frame &5e/pixel/frame \\
\hline
separation angle wrt zenith & 1~arcmin & 4/3~arcmin \\
\hline
science wavelength $\lambda$ & \multicolumn{2}{c}{$2.2~\mu m$} \\
\end{tabular}
\end{center}
\end{table}

To understand the influence of different guide stars on the PSF, the tests are performed with different photon flux, varying from 50 to 1000 photons per subaperture per frame for the LGS, but during one test run the flux does not change. The photon flux for the three NGS is fixed to 50000 photons per subaperture per frame.  Additionally, we simulate a WFS detector read-out noise. Each simulation run consists of 1000 time steps, corresponding to 2 seconds of real time. Note that we do not consider the effect of spot elongation for the LGS. \\

For our setting, OCTOPUS provides 25 true reference PSFs, calculated from the residual wavefront after DM correction on a regular $5 \times 5$ grid at $2.2~\mu m$ (K-band). In the reconstruction process, we only use directions for which a reference PSF is evaluated in order to be able to compare our reconstruction to a true PSF. The focus is on the center direction and on two off-axis directions, being $30''$ and $1'$ away from the center along the horizontal axis. Note that none of these three directions is a guide star direction, so we have to rely on atmospheric tomography. Furthermore, we chose these angles, since instruments of ESO's ELT will have a field of view of around $1'$, and LGS and NGS will be outside this field.

\subsection{Using the higher order components of the incoming wavefront}
Similar to PSF reconstruction for an SCAO system  in Ref.~\cite{WaHoRa18}, we need to estimate the higher order structure function $\bar{D}_{\phi_\perp}$ through the higher order components of the incoming phase. Note that we cannot simply reuse the structure function $\bar{D}_{\phi_\perp}$ from an SCAO simulation as the higher frequencies in the turbulence of the atmosphere corrected by the three DMs in MCAO and one DM in SCAO will not fully coincide. However, we can use the same method as in Ref.~\cite{WaHoRa18}: We simulate a large number of random phase screens using a Kolmogorov statistic for the specific atmospheric conditions. From these phase screens, we remove the modes which can be corrected by the DMs.
Then we downsample the residual phase to bigger pixel size, since otherwise calculating the structure function would be computationally too expensive.  In this downsampling procedure, the choice of the basis functions for numerical implementation plays a crucial role, as discussed in detail in Ref.~\cite{WaHoRa18}.
All results are obtained for a pixel size $\delta x = \frac{1}{4} d_{DM} = 0.125~m$, since $d_{DM} = 0.5~m$ is the spacing of the actuators of the ground layer DM in our simulations, and also the size of one WFS subaperture.\\

Additionally, in order to avoid temporal correlation in $\bar{D}_{\phi_\perp}$, we perform the temporal average not over every  $\phi_{\perp}$ but take only every 5th of these residual phases. Extensive test runs have shown that taking every 5th screen provides a good trade-off between computational requirements and resulting accuracy. This is also in good agreement with the turbulence coherence time for the minimal turbulence scale $\delta x =0.125~m$. Note that this temporal downsampling is done only for the computation of $\bar{D}_{\phi_\perp}$. \\

Furthermore note that the $c_n^2$-profile used for the simulation is not changing and will in reality not perfectly match with the actual one during observation.
OCTOPUS has a parameter called \emph{turbulent\_seed}, defining the starting point for the generation of atmospheric layers in a pseudo-random way. In order to avoid an unrealistic setting, we take different values of \emph{turbulent\_seed} for estimating $\bar{D}_{\phi_\perp}$ and the on-the-fly computation for estimating $D_{\phi_{\|,\beta_i}}$. This prevents in particular that the two structure functions $\bar{D}_{\phi_\perp}$ and $D_{\phi_{\|,\beta_i}}$ match perfectly for the considered atmosphere. However, using the same atmospheric profile is still a very optimistic approach.

\subsection{Numerical results for high photon flux}\label{sect:psfr-mcao-highflux-res}

\begin{figure}[!htb]
  \begin{center}
    \begin{tabular}{c}
\includegraphics[height=5cm]{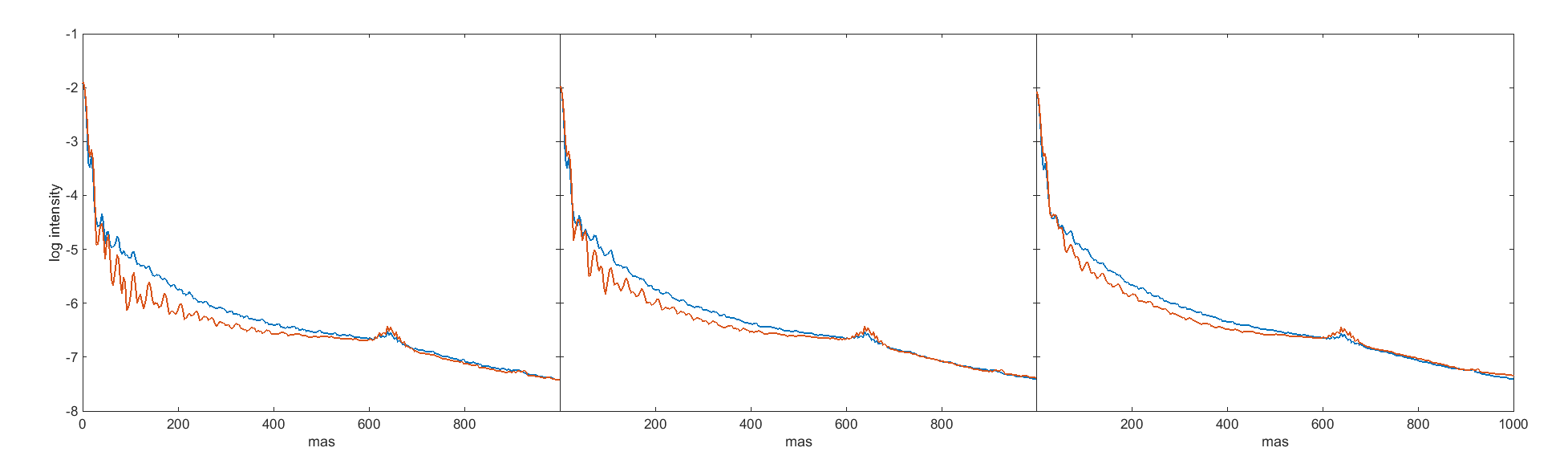}
  \end{tabular}
\end{center}
   \caption[PSF reconstruction for $n_{ph} = 1000$]{\label{fig:psfr-mcao1000}Comparison of the true PSF (blue) and the reconstructed PSF (red) for $n_{ph}=1000$ in center direction (left), at $30''$ (middle) and $60''$ (right) off-axis, radial average.}
 \end{figure}

In a first step, we analyze the high flux case with $n_{ph} = 1000$, i.e., 1000 photons reach each subaperture of an LGS WFS in every time step in the setting of Table~\ref{tab:mcao-system}. Since no guide star in center direction is available, the reconstructed PSF in center direction completely relies on the tomographic reconstruction. Figure~\ref{fig:psfr-mcao1000} shows the PSF for the center and two off-axis directions at $30''$ and $60''$, respectively, computed by OCTOPUS (blue) and the reconstructed PSF using a 4D structure function (red). All depicted PSFs are normalized to be energy preserving, i.e., $\|\mathcal{PSF}\|_1 = 1$. As a consequence, the Strehl ratio cannot be directly derived from these plots.  The PSFs are obtained in the pupil plane and not in the focal plane of the science camera. Note that we do not consider NCPAs and additional telescope specific effects (which are discussed, e.g., in Ref.~\cite{MICADO_PSFR20}) in our simulations. Furthermore, note that we only show the inner 1000~mas of the PSF. The part inside the control radius around 630~mas relates to the quality of our reconstruction of the parallel part structure function $D_{\phi_\|}$, while the part further outside relats to the orthogonal part structure function $D_{\phi_\perp}$ and thus is a statistical average corresponding to the current seeing.
%Inside the control radius around 630~mas we see the performance of our algorithm, while outside the control radius the structure of the PSF is a statistical average corresponding to the current seeing. 
We find a good agreement of the true and the reconstructed PSF in the center, leading to a mismatch in the peak by only $0.6\%$ using the 4D structure function. The Strehl ratio at $2.2~\mu m$ calculated by OCTOPUS for the true PSF in the center is $55.1\%$, where we used the Gradient-based method for AO control. \\

We reuse the higher order structure function $\bar{D}_{\phi_\perp}$ from the center direction for the off-axis directions. At a distance of $30''$, we find an underestimation of the peak in the reconstructed PSF by $1\%$, which is still within an acceptable range. Note that the Strehl ratio calculated by OCTOPUS for the true PSF in this direction is $48.8\%$, within the same simulation as used for the center direction. \\

Further outside, at $1'$ from the center, the Strehl ratio drops to $37.5\%$. Our method can still recover the Strehl ratio with an error of $1\%$ .
  We summarize all results in Table~\ref{tab:results-high}.
 
 \begin{table}[htbp]
\caption{Comparison of Strehl ratios at $2.2~\mu m$ for high photon flux}	
\label{tab:results-high}
\begin{center}
\begin{tabular}{l|c|c|c}
Off-axis distance & $0''$ & $30''$ & $60''$ \\
\hline
true Strehl ratio & 55.1\% & 48.8\% & 37.5\% \\
\hline
reconstructed Strehl ratio & 54.9\% & 48.3\% & 37.8\%  \\
\hline
error & 0.6\% &1\% & 1\% \\
\end{tabular}
\end{center}
\end{table}

\subsection{Numerical results for low photon flux}

Let us now consider a lower LGS photon flux, leading to a diminished AO performance. The telescope and simulation setting remain unchanged. Furthermore, the flux from the NGS is not altered.
We take the same $D_{\phi_\perp}$ as in the highflux case, which remains a reasonable estimate as the atmospheric and telescope parameters remain the same.\\% Similar to considering an off-axis direction in the previous section, this leads to an error in the estimation of $B_\perp$ and finally in $\mathcal{PSF}$.\\

\begin{figure}[!htb]
  \begin{center}
    \begin{tabular}{c}
\includegraphics[height=5cm]{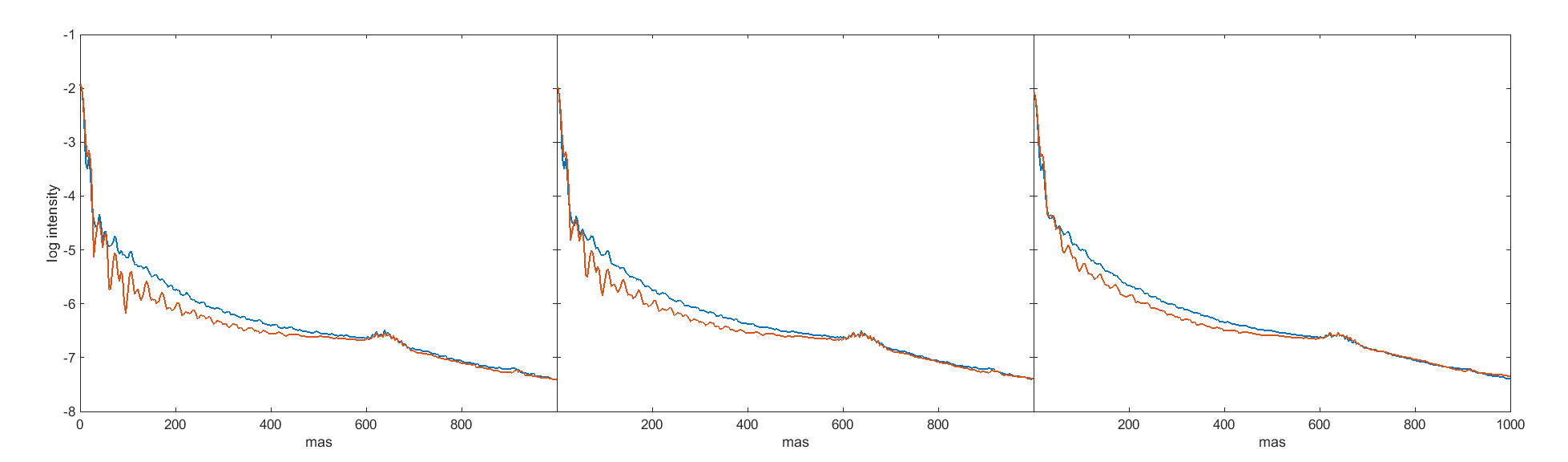}
  \end{tabular}
\end{center}
   \caption[PSF reconstruction for $n_{ph} = 500$]{\label{fig:psfr-mcao500}Comparison of the true PSF (blue) and the reconstructed PSF (red) for $n_{ph}=500$ in center direction (left), at $30''$ (middle) and $60''$ (right) off-axis, radial average.}
 \end{figure}

We start with $n_{ph} = 500$. The reconstructed PSF in center direction shows a good agreement with the true PSF, as illustrated in Figure~\ref{fig:psfr-mcao500}. The error in the peak is $1.7\%$. As in the previous section, for the AO control during our simulation run, we use the three-step approach with the gradient-based method leading to a Strehl ratio of $53.7\%$ in center direction.\\

Performing the same $30''$ off-axis estimation as for the high photon flux gives an underestimation of the peak by $2.9\%$, at a Strehl ratio of $47.6\%$ calculated by OCTOPUS. Similarly for $1'$ off-axis, we obtain an error of $3.8\%$ in the peak, at a Strehl ratio of $36.4\%$.\\

\begin{figure}
\begin{center}
\begin{tabular}{c}
\includegraphics[height=5cm]{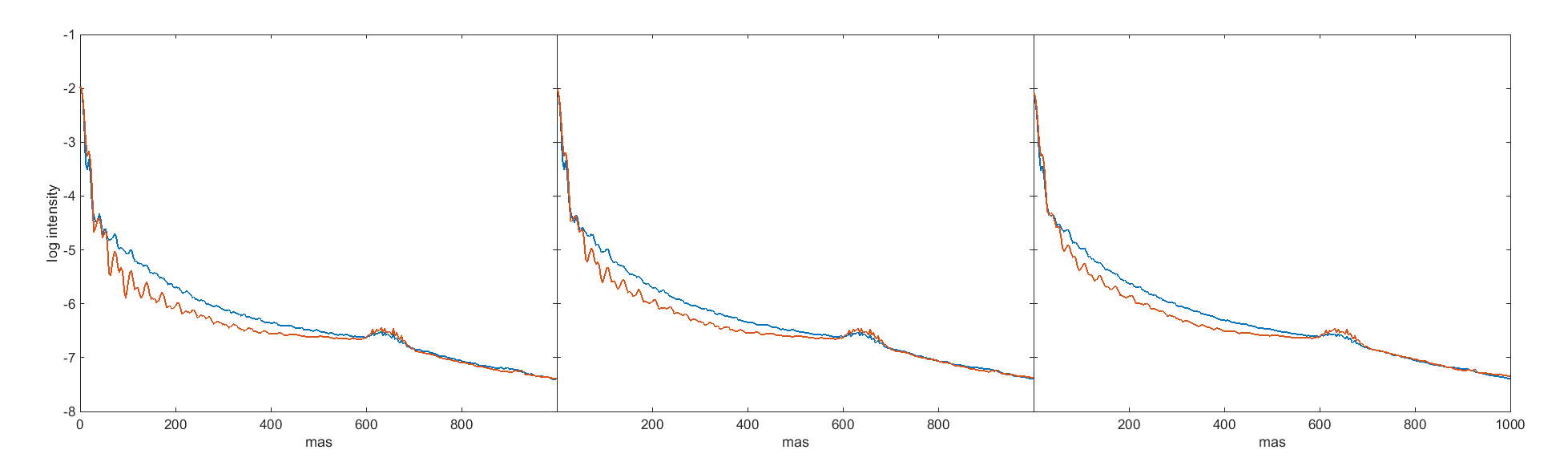}
\end{tabular}
\end{center}
   \caption[PSF reconstruction for $n_{ph} = 100$]{Comparison of the true PSF (blue) and the reconstructed PSF (red) for $n_{ph}=100$ in center direction (left), at $30''$ (middle) and $60''$ (right) off-axis, radial average.}
\label{fig:psfr-mcao100}
 \end{figure}

Lowering the photon flux from the LGS further to $n_{ph} = 100$ results in the PSFs shown in Figure~\ref{fig:psfr-mcao100} with a Strehl ratio of $47.8\%$ at the center. By coincidence, we estimate the peak perfectly. In the two off-axis directions, the Strehl ratio is $42.0\%$ and $31.6\%$ at $30''$ and $1'$, respectively. At $30''$ off-axis, we overestimate the peak by $2.2\%$. However, at this level, we overestimate the peak at $1'$ off-axis by $15.9\%$, indicating the limitation of our current algorithm using a standard Gaussian noise model only.\\

\begin{figure}
\begin{center}
\begin{tabular}{c}
\includegraphics[height=5cm]{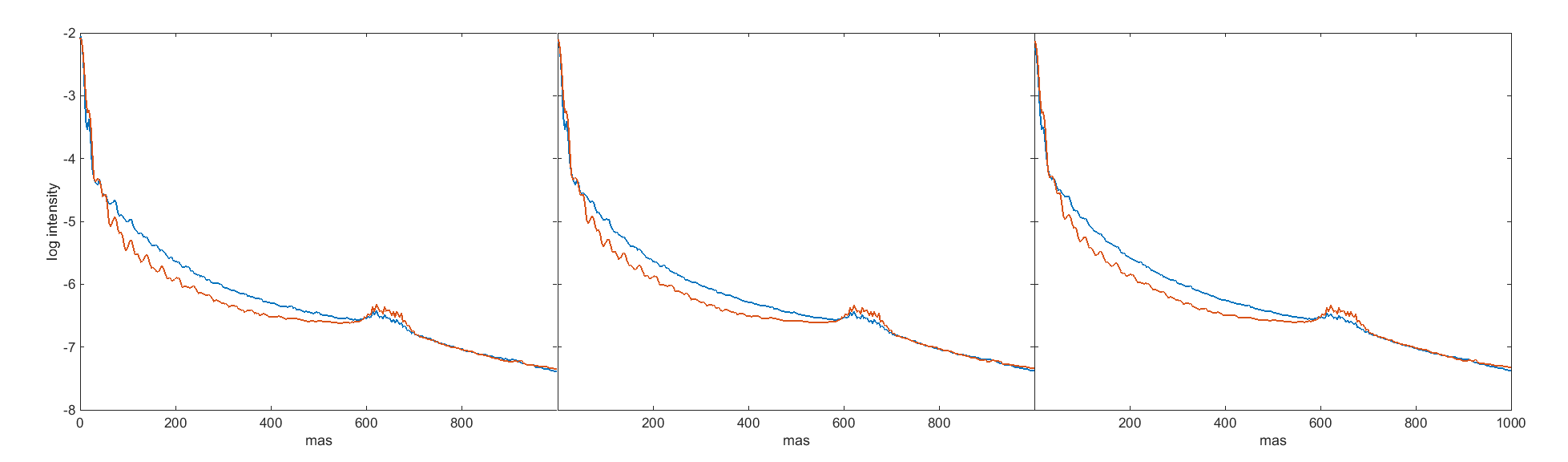}
\end{tabular}
\end{center}
\caption[PSF reconstruction for $n_{ph} = 50$]{\label{fig:psfr-mcao50}  Comparison of the true PSF (blue) and the reconstructed PSF (red) for $n_{ph}=50$ in center direction (left), at $30''$ (middle) and $60''$ (right) off-axis, radial average.}
\end{figure}

A further reduction of the LGS flux to $n_{ph} = 50$ leads to the PSFs shown in Figure~\ref{fig:psfr-mcao50}. We underestimate the Strehl ratio, being now $39.5\%$, by $3.5\%$ in center direction. In the two off-axis directions we have a Strehl ratio of $34.6\%$ and $25.6\%$ at $30''$ and $1'$, respectively. While the error is only $4\%$ at $30''$, our method gives a mismatch of $30\%$ at $1'$ off-axis. We summarize all results in Table~\ref{tab:results-low}.

 \begin{table}[htbp]
\caption{Comparison of Strehl ratios at $2.2~\mu m$ for low photon flux}	
\label{tab:results-low}
\begin{center}
\begin{tabular}{l|c|c|c|c|c|c|c|c|c}
LGS photon flux & \multicolumn{3}{c|}{500} & \multicolumn{3}{c|}{100} & \multicolumn{3}{c}{50}\\
\hline
Off-axis distance & $0'$ & $30''$ & $1'$ & $0'$ & $30''$ & $1'$ &$0'$ & $30''$ & $1'$ \\
\hline
true Strehl & 53.7\% & 47.6\% & 36.4\% & 47.8\% & 42.0\% & 31.6\% & 39.5\% & 34.6\% & 25.6\% \\
\hline
reconstructed Strehl & 52.8\% & 46.2\% & 37.8\% & 47.8\% & 41.1\% & 36.6\% & 38.1\% & 33.2\% & 33.2\%  \\
\hline
error & 1.7\% & 2.9\% & 3.8\% & 0\% & 2.2\% & 15.9\% & 3.5\% & 4\% & 30\% \\
\end{tabular}
\end{center}
\end{table}

In summary, we find that the reconstructed PSF matches the true PSF very accurately inside the region covered by the LGS. As expected, outside this area the reconstruction quality decreases, but our algorithm still provides reasonably accurate results for bright LGS.

\section{Conclusion and outlook}
In this work, we presented a new algorithm for PSF reconstruction in an MCAO system for the upcoming generation of GSMTs. Our approach uses reconstructed atmospheric layers to overcome the problem of field dependent PSFs in wide field AO systems. With these layers projected in the direction of interest, it is then possible to compute PSFs for different view directions within the field of view. First simulations show a qualitatively good reconstruction of the PSF compared to the PSF calculated directly from the simulated incoming wavefront.\\% Furthermore, the used algorithm has a reasonable run time and memory consumption.\\

The algorithm could further be improved by a more accurate model for the noise covariance used in the computations of the parallel part structure function $D_{\phi_\|}$ and by using telescope calibrations such as the ones proposed for MICADO\cite{MICADO_PSFR20}. A future goal of our work is to include this algorithm into the data reduction pipeline of MICADO in order to provide reconstructed PSFs once MAORY is attached to MICADO.
Furthermore, we aim to use the reconstructed PSFs as input to a blind deconvolution algorithm for image improvement, which can be done after the observation on the telescope only, e.g., Refs.~\cite{DyRaReSoWa21,RaSoHu21}. Such an approach leads to a further improvement of the quality of the reconstructed PSF and simultaneously improves the quality of the observed image.

\section*{Funding}
This work was funded by the Hochschulraumstrukturfonds of Austrian Ministry of research (BMWFW), project ``Beobachtungsorientierte Astrophysik in der E-ELT \"Ara'' and the Austrian Science Fund (FWF), project F6805-N36, SFB Tomography Across the Scales.

\printbibliography
\end{document}